\documentclass[prb,aps,twocolumn,showpacs]{revtex4}
\usepackage{amsfonts,amsthm}
\usepackage{amsmath}
\usepackage{amssymb}
\usepackage{graphicx}
\usepackage{dcolumn}
\usepackage{bm}
\usepackage{color}
\usepackage{ulem}

\begin{document}

\preprint{APS/123-QED}

\title{3D Dirac Electrons on a Cubic Lattice with Noncoplanar Multiple-$Q$ Order}

\author{Satoru Hayami}
 \author{Takahiro Misawa}%
 \author{Youhei Yamaji}%
\author{Yukitoshi Motome} 
\affiliation{%
 Department of Applied Physics, University of Tokyo, Tokyo 113-8656, Japan
}%

\begin{abstract}
Noncollinear and noncoplanar spin textures in solids manifest themselves not only in their peculiar magnetism but also in unusual electronic and transport properties. 
We here report our theoretical studies  
of a noncoplanar order on a simple cubic lattice and its influence on the electronic structure. 
We show that 
a four-sublattice triple-$Q$ order 
induces three-dimensional massless Dirac electrons at
commensurate electron fillings. 
The Dirac state is doubly degenerate, while it splits into
a pair of Weyl nodes by lifting the degeneracy by an external magnetic field; the system is turned into a Weyl semimetal in applied field. 
In addition, we point out the triple-$Q$ Hamiltonian in the strong coupling limit is equivalent to the 3D $\pi$-flux model relevant to an AIII topological insulator. 
We examine the stability of such a triple-$Q$ order in two fundamental models for correlated electron systems: 
a Kondo lattice model with classical localized spins 
and a periodic Anderson model. 
For the Kondo lattice model, performing a variational calculation and Monte Carlo simulation, 
we show that the triple-$Q$ order is widely stabilized around 1/4 filling.
For the periodic Anderson model, we also show the stability of the same triple-$Q$ state 
by using the mean-field approximation. 
For both models, 
the triple-$Q$ order is widely stabilized  
via the couplings between conduction electrons and localized electrons
even without any explicit competing magnetic interactions and geometrical frustration.  
We also show that the Dirac electrons induce peculiar surface states: Fermi ``arcs" connecting
the projected Dirac points, similarly to Weyl semimetals. 
\end{abstract}
\maketitle

\section{Introduction}
Noncoplanar magnetic orders, in which spin directions align neither in a line nor on a plane,
often lead to new low-energy excitations and topologically nontrivial states.
In particular, triple-$Q$ magnetic orders,
which are characterized by three different ordering
wave vectors, have drawn much interest. 
A skyrmion lattice, found, e.g., in the A phase of MnSi~\cite{Muhlbauer_2009skyrmion},  
is a typical example of such triple-$Q$ orders. 
In this case, the triple-$Q$ order is 
stabilized by competition between ferromagnetic exchange interaction and Dzyaloshinskii-Moriya interaction. 
Another example is 
found in geometrically frustrated lattices, 
which gives rise to a topological (Chern)
insulator 
and associated quantum anomalous Hall effect: for instance, on kagome~\cite{Ohgushi_PhysRevB.62.R6065}, 
distorted face-centered-cubic~(FCC)~\cite{Shindou_PhysRevLett.87.116801},
and triangular lattices~\cite{Martin_PhysRevLett.101.156402,Akagi_JPSJ.79.083711,Akagi_PhysRevLett.108.096401}.

In the present study,
we investigate how
a triple-$Q$ magnetic order
affects the single-particle spectrum of conduction electrons 
on a simple cubic lattice. 
By deriving the low-energy effective Hamiltonian, 
we reveal that a triple-$Q$ magnetic order generally accommodates
three-dimensional (3D) massless Dirac electrons on the cubic lattice.
Furthermore, we show that the triple-$Q$ magnetic order is widely stabilized 
in the weak-to-intermediate coupling region in the Kondo lattice model with classical localized spins. 
A similar triple-$Q$ state was obtained in the strong coupling limit 
as a consequence of the competition 
between the double-exchange ferromagnetic interaction 
and superexchange antiferromagnetic interaction~\cite{PhysRevB.64.054408}. 
In contrast, the present study reveals that
the triple-$Q$ state emerges even without such competition.
We also show that such a Dirac electronic 
state in a triple-$Q$ magnetic order is 
realized in 
the periodic Anderson model on the cubic lattice, 
which is a more generic model relevant for describing real materials such as heavy fermion systems. 

We also unveil 
peculiar properties of the 3D massless Dirac electrons associated with the triple-$Q$ order, which was not studied in detail previously~\cite{PhysRevB.64.054408}. 
One is the emergence of Weyl electrons in applied magnetic field. 
Our Dirac state is, at least, doubly degenerate. 
The degeneracy is lifted by applying magnetic field 
without gap opening, and the Dirac electronic state splits 
into a pair of Weyl states. 
Weyl electrons were recently proposed 
for an iridium pyrochlore oxide 
Y$_{2}$Ir$_{2}$O$_{7}$~\cite{Wan_PhysRevB.83.205101}. 
Our result offers yet another example of Weyl semimetals. 
Another interesting property is the emergence of surface states. 
Even without lifting the degeneracy of the Dirac electrons, 
our triple-$Q$ state exhibits peculiar gapless surface states with Fermi ``arcs",
similarly to 
Weyl semimetals~\cite{Wan_PhysRevB.83.205101}. 

Our finding 
may open new avenues for engineering massless Dirac electrons. 
Dirac electrons in a bulk material are classified into several categories: e.g., 
symmetry-protected 
ones~\cite{Dresselhaus_Dresselhaus_jorio,PhysRevB.85.155118,PhysRevB.85.195320,Young2012} as anticipated in graphene,
ones appearing only 
at the band-inversion transition points between
topologically trivial and nontrivial phases~\cite{Science.314.1757,Hasan_RevModPhys.82.3045}, 
and ones coexisting with spontaneous symmetry breaking, such as charge order (CO) in 
$\alpha$-(BEDT-TTF)$_2$I$_3$~\cite{Katayama_JPSJ.75.054705} and magnetic order in BaFe$_{2}$As$_{2}$~\cite{BaFe2As2_PhysRevLett.104.137001}. 
Among them, the symmetry-protected Dirac electrons 
are interesting from the viewpoint of potential 
applications for electronics and spintronics, 
as they are stable against perturbations 
which preserve the symmetry of the system. 
They, however, appear only in a limited number of crystalline lattice structures
because of severe constraints from the space group symmetry~\cite{PhysRevB.85.155118,Young2012}. 
Our result brings a new member of 
symmetry-protected massless Dirac electrons. 
This indicates that multiple-$Q$ orders exploit 
the possibility of engineering Dirac electrons 
by relaxing the symmetry constraints~\cite{PhysRevB.85.155118,Young2012}. 
Furthermore, our results on the Weyl states in applied magnetic field demonstrate the contollability of Dirac electrons via spin degree of freedom, which is potentially useful for spintronics.

The organization of this paper is as follows:
In Sec.~\ref{sec:3D_Dirac}, we show how the triple-$Q$
magnetic order induces the 3D massless Dirac electrons.
We present the detailed analysis of the low-energy effective Hamiltonian.  
We also point out that an external magnetic field splits the twofold degeneracy of the Dirac nodes and produces a pair of Weyl nodes. 
In Sec.~\ref{sec:stability}, we examine the stability of
the triple-$Q$ magnetic order in the 
Kondo lattice model and the periodic Anderson model.
For the Kondo lattice model with classical localized spins, by performing the variational calculation 
at zero temperature and Monte Carlo simulation for finite temperature,
we clarify that the triple-$Q$ magnetic order is indeed realized around 1/4 filling in the weak-to-intermediate coupling region.
For the periodic Anderson model, by using the mean-field approximation, 
we show that the triple-$Q$ magnetic order
appears in a wide range of phase diagram at zero temperature and at 3/4 filling.
In Sec.~\ref{sec:electronic_structure},
we examine the peculiar surface electronic structures in the triple-$Q$ state, by taking the results 
in the periodic Anderson model. 
Section~\ref{sec:summary} is devoted to summary and concluding remarks. 

\begin{figure}[htb!]
\begin{center}
\includegraphics[width=8cm,clip]{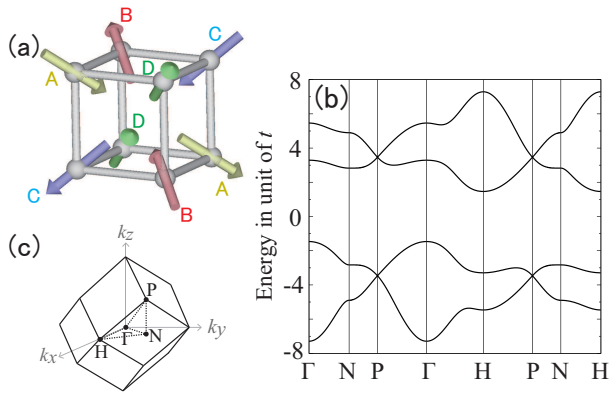} 
\caption{
(Color online).
(a) 
Schematic picture of the noncoplanar four-sublattice triple-$Q$ order on a cubic lattice. 
Each spin points along the local $[111]$ directions. 
A global spin rotation does not alter the consequences, as the system is SU(2) symmetric. 
A-D denote the four sublattices, respectively. 
(b)
Energy dispersion of the Hamiltonian in Eq.~(\ref{eq:H_reduced}) at $\Delta=2t$, 
shown along the symmetric lines in the magnetic Brillouin zone displayed in (c), by connecting 
$\Gamma = (0, 0, 0)$, $N = (\pi/2, \pi/2, 0)$, $P = (\pi/2, \pi/2, \pi/2)$, and $H = (\pi, 0, 0)$. 
3D massless Dirac points appear at the $P$ point, corresponding to 1/4 and 3/4 fillings. 
\label{Fig:ponti_Dirac_dispersion}
}
\end{center}
\end{figure}

\section{3D massless Dirac electrons}
\label{sec:3D_Dirac}

Let us begin with explaining how a 
triple-$Q$ magnetic order induces 3D massless Dirac electrons. 
We consider noninteracting 
electrons locally coupled to a triple-$Q$ magnetic order on the cubic lattice set by spins 
\begin{eqnarray} 
\bm{S}_{i} \propto [\cos(\bm{Q}_1\cdot \bm{r}_{i}),\cos(\bm{Q}_2\cdot \bm{r}_{i}),\cos(\bm{Q}_3 \cdot \bm{r}_{i})]. 
\label{eq:tripleQ}
\end{eqnarray}
Here, 
$\bm{r}_i$ is the position vector of the site $i$ on the cubic lattice with the lattice constant $a=1$; 
$\bm{Q}_1=(\pi ,0, \pi)$, 
$\bm{Q}_2=(0, \pi ,\pi)$, 
$\bm{Q}_3=(\pi ,\pi, 0)$ 
represent the wave vectors characterizing the triple-$Q$ state. 
$\bm{S}_i$ has a noncoplanar four-sublattice structure in the real space, as 
schematically shown in Fig.~\ref{Fig:ponti_Dirac_dispersion}(a). 
The Hamiltonian reads 
\begin{eqnarray}
{\mathcal{H} }  = \sum_{  \bm{k}, \sigma}  \epsilon_{\bm{k}}
c^{\dagger}_{\bm{k} \sigma} c_{\bm{k} \sigma}  
+ \Delta \sum_{\bm{k} ,\sigma, \sigma', \eta} 
c^{\dagger}_{\bm{k} \sigma} \bm{\sigma}^{{\sigma\sigma'}}_{{\eta}} c_{\bm{k}+\bm{Q}_{\eta} \sigma^{\prime}}, 
\label{Kondo_Ham_kspace}
\end{eqnarray} 
where $c^{\dagger}_{\bm{k} \sigma}$($c_{\bm{k} \sigma}$) 
is the creation (annihilation) operator of a conduction electron 
with spin $\sigma$ at wave vector $\bm{k}$. 
The first term represents the 
kinetic energy of conduction electrons and $\epsilon_{\bm{k}}$ is the 
energy dispersion for free electrons 
on the cubic lattice, $\epsilon_{\bm{k}} = -2t (\cos k_x +\cos k_y +\cos k_z)$.  
The second term describes the coupling to triple-$Q$ 
magnetic order, where $\Delta=J/(2\sqrt{3})$ when we take $|\bm{S}_i|=1$; $J$ is the coupling constant between the conduction electron spins and $\bm{S}_i$, and $\bm{\sigma}$ is the Pauli matrix ($\eta=1$, $2$, and $3$) [see also Eq.~(\ref{Kondo_Ham_RS})]. 
In the four-sublattice representation, 
the Hamiltonian is divided into two irreducible parts
\begin{eqnarray}
{\mathcal{H} }  = \bm{c}_{I}^{\dagger} {\tilde{\mathcal{H}}} \bm{c}_{I} + \bm{c}_{I \!I}^{\dagger} {\tilde{\mathcal{H}}} \bm{c}_{I\!I}, 
\label{eq:H_reduced}
\end{eqnarray}
where 
\begin{equation}
{\tilde{\mathcal{H}}} =\left( \begin{array}{cccc}
\epsilon_{\bm{k}} & \Delta & -{\rm i}\Delta &\Delta\\
\Delta & \epsilon_{\bm{k}+\bm{Q}_1} & -\Delta & {\rm i}\Delta \\
{\rm i} \Delta & -\Delta & \epsilon_{\bm{k}+\bm{Q}_2} & \Delta \\
\Delta & -{\rm i}\Delta & \Delta & \epsilon_{\bm{k}+\bm{Q}_3} 
\end{array} \right). 
\label{effective_Hamiltonian}
\end{equation}
Here, $\bm{c}_{I}^{\dagger} =$ ($c_{\bm{k}\uparrow}^{\dagger}$, $c_{\bm{k}+\bm{Q}_1 \downarrow}^{\dagger}$, $c_{\bm{k}+\bm{Q}_2 \downarrow}^{\dagger}$, $c_{\bm{k}+\bm{Q}_3 \uparrow}^{\dagger}$) and 
$\bm{c}_{I\!I}^{\dagger} =$ ($c_{\bm{k}\downarrow}^{\dagger}$, $c_{\bm{k}+\bm{Q}_1 \uparrow}^{\dagger}$, $-c_{\bm{k}+\bm{Q}_2 \uparrow}^{\dagger}$, $-c_{\bm{k}+\bm{Q}_3 \downarrow}^{\dagger}$). 
We note that this Hamiltonian is 
formally the same as that for the four-sublattice triple-$Q$ order on 
the triangular lattice~\cite{Martin_PhysRevLett.101.156402}. 
In the triangular lattice, the triple-$Q$ magnetic order induces the full gap 
while in the cubic lattice the Dirac dispersions remain as we show below. 

The energy dispersion of the Hamiltonian in Eq.~(\ref{eq:H_reduced}) is shown 
in Fig.~\ref{Fig:ponti_Dirac_dispersion}(b) at $\Delta=2t$.
Here, all the bands are 
doubly degenerate; 
the degeneracy comes from the fact that $\bm{c}_{I}$ and $\bm{c}_{I\!I}$ are 
related by a combination of lattice translation and 
spin rotation, which leaves ${\tilde{\mathcal{H}}}$ unchanged. 
In Fig.~\ref{Fig:ponti_Dirac_dispersion}(b), 
a peculiar form of dispersions is found near the $P$ point, 
i.e., $\bm{k} \simeq (\pi/2, \pi/2, \pi/2)$; 
the band dispersions are linearly dependent on $\bm{k}$ and cross with each other at the $P$ point, 
resulting in 3D cone-like structures. 
This is a signature of 3D massless Dirac electrons appearing at 1/4 and 3/4 fillings of electrons. 

The Dirac-type dispersion indeed follows the Dirac equation as follows. 
This is explicitly shown 
by deriving a low-energy Hamiltonian near the $P$ point by the $\boldsymbol{k}\cdot\boldsymbol{p}$ perturbation theory~\cite{Dresselhaus_Dresselhaus_jorio}. 
Expanding the reduced Hamiltonian in Eq.~(\ref{effective_Hamiltonian}) around the $P$ point and 
performing the unitary transformations, we obtain the 
low-energy effective Hamiltonian 
up to the first order in $t|\bm{\kappa}|/ \Delta$ as 
\begin{eqnarray}
{\tilde{\mathcal{H}}}_{\pm}=\pm \sqrt{3} \Delta {\sigma_{0}} \pm \frac{2}{\sqrt{3}} t (\kappa_x {\sigma_{3}} + \kappa_y {\sigma_{1}} + \kappa_z {\sigma_{2}}), 
\label{effective_Weyl_Hamiltonian2}
\end{eqnarray}
where $\sigma_0$ is the $2\times 2$ identity matrix and 
$\bm{\kappa}$ is the transformed wave vector measured from the $P$ point (see Appendix~\ref{Low-energy Hamiltonian in the triple-Q phase} for the derivation). 
This Hamiltonian constitutes a set of 
four-component Dirac Hamiltonian $\mathcal{H}_{\pm}^{\rm eff}=\tilde{\mathcal{H}}_{\pm}\otimes \mbox{\boldmath$I$}_{2\times 2}$
that describes the 3D massless Dirac electrons with a linear dispersion in all the three directions of
$\bm{\kappa}$. 

The four-component Dirac electrons are not chiral,
as there is no unitary matrix which anticommutes with the low-energy Hamiltonian. 
Such a non-chiral 3D massless Dirac state cannot be turned into an AIII topological insulator~\cite{ryu2010topological} 
by opening a gap, at least, within four-sublattice unit cell.  
As clarified in Ref.~\onlinecite{PhysRevB.81.045120}, however, 
a 3D $\pi$-flux model can change into the AIII topological insulator by an appropriate perturbation, which has eight-sublattice structure.  
As detailed in Appendix~\ref{Appendix2}, 
we notice that the triple-$Q$ Hamiltonian in Eq.~(\ref{Kondo_Ham_kspace}) in the strong coupling limit 
$\Delta/t \rightarrow \infty$ is indeed equivalent
to the 3D $\pi$-flux model studied in Ref.~\onlinecite{PhysRevB.81.045120}. 
Thus, our triple-$Q$ state can be also switched into the AIII topological insulator by an appropriate perturbation. 

As we mentioned above, the Dirac states in Eqs.~(\ref{eq:H_reduced}) 
and (\ref{effective_Weyl_Hamiltonian2}) are doubly degenerate. 
We, however, find that the twofold degeneracy is lifted by an external magnetic field. 
By adding the Zeeman term under magnetic field $H$ applied in the $z$ 
direction 
\begin{align}
\mathcal{H}_{\rm Z} = -H \sum_{\bm{k}} (c_{\bm{k} \uparrow}^{\dagger}c_{\bm{k} \uparrow} - c_{\bm{k} \downarrow}^{\dagger} c_{\bm{k} \downarrow}), 
\end{align}
the degenerate Dirac point is split into two, and they are shifted 
to the opposite directions along the $k_z$ axis. 
The resultant nondegenerate nodes accommodate Weyl electrons, and the Fermi level is pinned at the nodes. 
The system, therefore, is turned into a Weyl semimetal by applied magnetic field. 
The Weyl state is robust against any perturbations that preserve 
the symmetry of the system. 
Thus, our model gives an example of Weyl semimetals on an unfrustrated lattice, 
distinct from those on a frustrated pyrochlore lattice~\cite{Wan_PhysRevB.83.205101}. 

\section{Stability of Triple-$Q$ phase}
\label{sec:stability}
In the previous section, we simply assumed the noncoplanar 
triple-$Q$ magnetic order and discussed the resultant electronic state. 
Now, we examine when and how the triple-$Q$ state is realized. 
We here consider two fundamental models for $d$- and $f$-electron 
compounds, the Kondo lattice model (Sec.~\ref{Kondo lattice model}) and the periodic Anderson model (Sec.~\ref{sec:PAM}) on the cubic lattice. 

\subsection{Kondo lattice model}
\label{Kondo lattice model}

The Kondo lattice Hamiltonian is written by
\begin{align}
\label{Kondo_Ham_RS}
{\mathcal{H}}   
= -  t \sum_{\langle i,j\rangle,\sigma} 
( c^{\dagger}_{i \sigma} c_{j \sigma} +{\rm H.c.}) + \frac{J}{2} \sum_{\bm{i} ,\sigma, \sigma'} 
c^{\dagger}_{i \sigma} \bm{\sigma}^{{\sigma\sigma'}} c_{i \sigma^{\prime}} \cdot \bm{S}_{i}, 
\end{align}
where $\bm{S}_{i}$ is a localized moment. 
Here, we assume $\bm{S}_{i}$ to be a classical spin with $|\bm{S}_{i}|=1$. 
The first term represents 
the kinetic energy of conduction electrons 
and the second term represents 
the onsite interaction between conduction and localized spins. 
The sign of the coupling constant $J$ is irrelevant for the classical spins. 
For a fixed triple-$Q$ spin configuration given in Eq.~(\ref{eq:tripleQ}), the model is reduced to that in Eq.~(\ref{Kondo_Ham_kspace}). 
Hereafter, we take $t=1$ as an energy unit. 
We here examine 
the stability of the triple-$Q$ state with 3D Dirac electrons 
in the ground state in Sec.~\ref{Variational study for the ground state} and at finite 
temperature in Sec.~\ref{Monte Carlo study at finite temperature}. 

\subsubsection{Variational study for the ground state}
\label{Variational study for the ground state}
\begin{figure}[htb!]
\begin{center}
\includegraphics[width=8cm,clip]{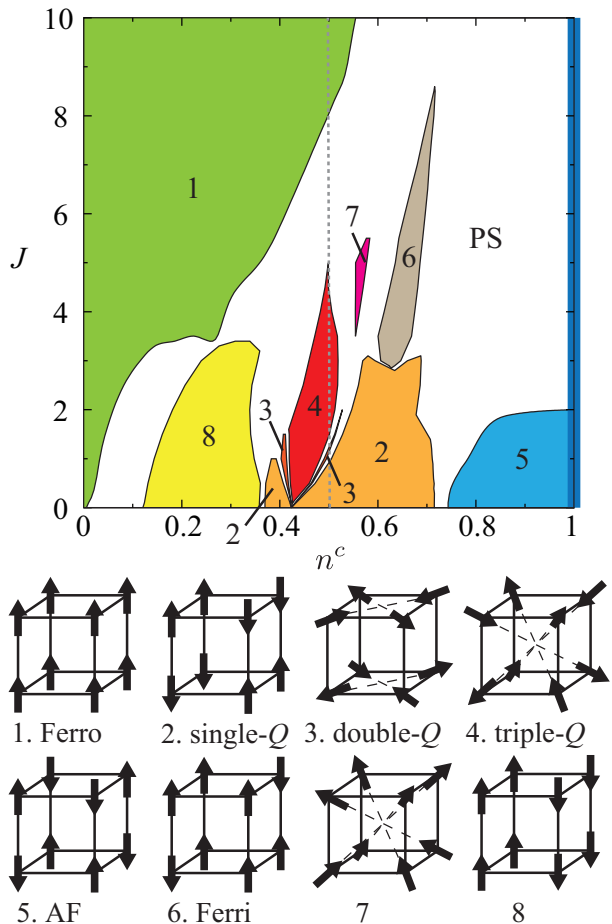} 
\caption{
(Color online). Ground-state phase diagram of the Kondo 
lattice model [Eq.~(\ref{Kondo_Ham_RS})] by the variational calculations. 
The vertical dashed line denotes $n^c=0.5$. 
PS indicates a phase separated region.  
Ordering patterns of localized spins are shown in the bottom of panel. 
Ferro, AF, and Ferri stand for ferromagnetic, antiferromagnetic, and ferrimagnetic, respectively. 
Single-$Q$ corresponds to $\bm{Q} = (\pi, 0, \pi)$, 
double-$Q$ $(0, \pi, \pi)$, $(\pi, 0, \pi)$, and
triple-$Q$ $(\pi, \pi, 0)$, $(0, \pi, \pi)$, $(\pi, 0, \pi)$. 
}
\label{Fig:Phase_diagram_KLM}
\end{center}
\end{figure}

First, we examine the ground state of the 
model given by Eq.~(\ref{Kondo_Ham_RS}) by 
changing $J$ and the electron density 
$n^c =(1/N)\sum_{i \sigma} \langle c_{i \sigma}^{\dagger}c_{i \sigma}  \rangle$ 
where $N$ is the total number of sites. 
We perform a variational calculation: we compare the zero-temperature grand potential 
$\Omega = E-\mu n^c$ ($E=\langle{\mathcal{H}}\rangle/N$ is the internal energy per site and $\mu$ is the chemical potential) 
for different ordered states of the localized spins and 
determine the most stable ordering. 
We assumed a collection of collinear and coplanar spin states, and found that the eight of them give the ground state in the range of parameters we studied (see Fig.~\ref{Fig:Phase_diagram_KLM}). 
In this calculation, we consider only uniform $\bm{q}=0$ orders for all the spin patterns. 
Note that unbiased Monte Carlo calculations do not show any signatures 
of longer-period orders around the most interesting 1/4 filling, as we detail later in Sec.~\ref{Monte Carlo study at finite temperature}.

Figure~\ref{Fig:Phase_diagram_KLM} shows the result of the phase diagram as a function of $n^c$ and $J$. 
In the low-density region, the ferromagnetic metallic phase appears and becomes wider as $J$ increases. 
This ferromagnetic phase is stabilized by the double-exchange mechanism~\cite{PhysRev.82.403} 
($J$ is antiferromagnetic, but the sign is irrelevant in the current study, as mentioned above). 
In contrast to this, 
a N\'eel-type antiferromagnetic order emerges at and near half-filling ($n^c=1$). 
This is partly understood by considering 
the second-order perturbation 
with respect to $t/J$ at half filling, which leads 
to an effective antiferromagnetic interaction between localized spins. In the weak-coupling region apart from half-filling, however,  
an incommensurate order might 
take over when taking account of longer-period orders. 

For intermediate $n^c$, the phase diagram becomes more complicated. 
Among various phases, we find that 
a four-sublattice triple-$Q$ order 
is realized near 1/4 filling ($n^c=0.5$), as shown in Fig.~\ref{Fig:Phase_diagram_KLM}. 
The band structure in this triple-$Q$ phase has the 3D Dirac nodes, equivalent to that in Fig.~\ref{Fig:ponti_Dirac_dispersion}(c). 
In this case, however, even at 1/4 filling, the Dirac nodes are located slightly above the Fermi level 
and an electron pocket appears at the $H$ point. 
The Fermi level comes at the Dirac nodes for $J>6$, where 
the triple-$Q$ magnetic order is preempted by the phase separation
or ferromagnetic order, as shown in Fig.~\ref{Fig:Phase_diagram_KLM}.
 
The stable Dirac electrons at the Fermi level for $n^c=0.5$ 
are realized by introducing other interactions. 
In fact, the triple-$Q$ phase becomes much more stable 
by including the small 
next-nearest-neighbor interaction between conduction and localized spins 
and/or the nearest-neighbor interaction between localized spins (not shown). 
In the latter connection, we note that a similar triple-$Q$ state was obtained in the strong coupling limit $J/t \to \infty$ 
in the presence of an antiferromagnetic interaction between neighboring localized spins~\cite{PhysRevB.64.054408}, 
although our triple-$Q$ order is stabilized even without any explicit competing interactions. 

\subsubsection{Monte Carlo study at finite temperature}
\label{Monte Carlo study at finite temperature}

\begin{figure}[htb!]
\begin{center}
\includegraphics[width=8cm,clip]{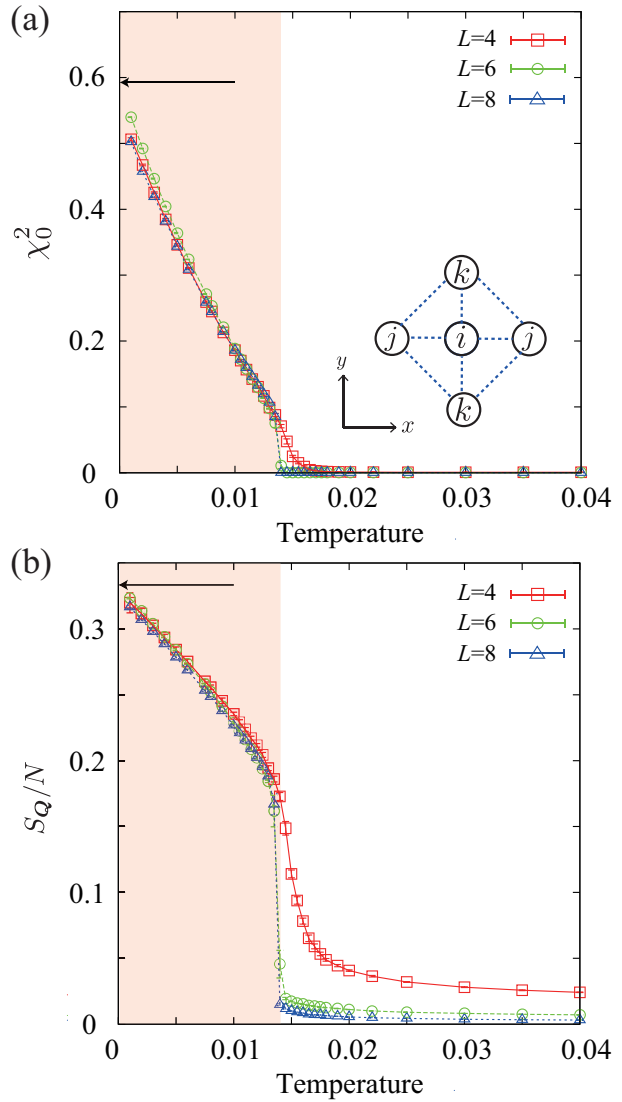} 
\caption{
Temperature dependence of (a) the square of local spin scalar chirality and (b) the spin structure factor averaged over ${\bm q} = {\bm Q}_1$, ${\bm Q}_2$, and ${\bm Q}_3$ divided by the system size $N$ at $J=4$. The arrows 
represent the values in the perfectly triple-$Q$ ordered state.
The inset in (a) shows the relation among sites $i$, $j$, and $k$ in Eq.~(\ref{eq:chi_0}). 
}
\label{Fig:chiral_sq}
\end{center} 
\end{figure}

Next, to examine whether the triple-$Q$ order 
is stable against 
spatial and thermal fluctuations, we perform 
Monte Carlo simulation for the Kondo lattice model,
which is a numerically exact method within statistical errors. 
In the Monte Carlo calculations, we first obtain the eigenstates for conduction 
electrons 
for given spin configurations by diagonalizing the Hamiltonian in Eq.~(\ref{Kondo_Ham_RS}). 
Then, by using the eigenvalues, we update the local spins according to the standard Metropolis method. 
Note that the simulation does not suffer from the negative-sign problem. 
The calculations were typically done for $8000$-$20000$ Monte Carlo steps and the statistical errors were estimated by dividing the data into ten bins. 
The calculations were conducted on the $N(=L^3)$-site cubic lattice with $L$ = 4, 6, and 8
under the periodic boundary conditions. 
For $L=4$ and $L=6$, we introduce supercells consisting of $N_{\bm{k}} = 2^3$ copies of the
$N$-site cube under periodic boundary conditions. The introduction of the supercells reduces
the finite size effects in the simulations. 
We take the Boltzmann constant $k_{\rm B}=1$. 

We here calculate temperature dependence of 
two physical quantities to identify the triple-$Q$ state. 
One is the local spin scalar chirality $\chi_0$ and the 
other is the spin structure factor $S_{{\bm q}}$. 
The local spin scalar chirality is defined by
\begin{eqnarray}
\chi_{0} &=& \frac{1}{2 zN}\sum_{i} \sum_{ \{j,k\} \in \Gamma_{i}} \langle  {\bm S}_{i} \cdot {\bm S}_{j} \times {\bm S}_{k} \rangle, 
\label{eq:chi_0}
\end{eqnarray}
where $z=6$ is the number of nearest-neighbor sites. 
The sum of $i$ is taken over all the sites, and 
$\Gamma_{i}$ represents the set of sites 
$j,k$ 
defined as follows: 
$j$th site represents a site next to $i$th site along the $x$ $(y, z)$ 
direction and $k$th site represents a site next to $i$th site 
along the $y$ $(z, x)$ direction in the $xy$ $(yz, zx)$ plane, 
as exemplified in the inset of Fig.~\ref{Fig:chiral_sq}(a). 
The spin structure factor is defined by 
\begin{eqnarray}
S_{{\bm q}} = \frac{1}{N} \sum_{i,j} \langle \bm{S}_{i} 
\cdot \bm{S}_{j} \rangle e^{i \bm{q} \cdot (\bm{r}_i - \bm{r}_j)}, 
\end{eqnarray}
where ${\bm q}$ is a wave vector. 
For the perfectly triple-$Q$ ordered state, 
the square of the local spin scalar chirality, $\chi_{0}^2$, is $\simeq0.59$, and the 
spin structure factor $S_{{\bm q}}$ takes the same value $\simeq 0.33 N$ at the triple-$Q$ wave numbers, ${\bm q} = {\bm Q}_1$, ${\bm Q}_2$, and ${\bm Q}_3$. 

Figure~\ref{Fig:chiral_sq} shows the Monte Carlo results. 
We here display $\chi_0^2$ in Fig.~\ref{Fig:chiral_sq}(a) and the averaged spin structure factor, $S_{\bm Q} = (S_{{\bm Q}_1}+S_{{\bm Q}_2}+S_{{\bm Q}_3})/3$ divided by $N$ in Fig.~\ref{Fig:chiral_sq}(b). 
As shown in the figures, 
the transition from the paramagnetic state to 
the triple-$Q$ state occurs at $T \simeq 0.014$ as decreasing temperature; 
the chiral order and magnetic order occur simultaneously~\cite{comment_Sq}.  
This is in contrast to the two-dimensional triangular lattice case, where the chiral order occurs alone at a finite temperature~\cite{PhysRevLett.105.266405}. 
The concomitant transition in our model appears to be of first order. 
With further decreasing temperature, the local spin scalar chirality and spin 
structure factor approach their saturated values for the triple-$Q$ state in the ground state. 
The results clearly show that the triple-$Q$ state found in the variational calculations remains stable against spatial and 
thermal fluctuations.

\subsection{Periodic Anderson model}
\label{sec:PAM}

Next, we examine the stability of noncoplanar triple-$Q$ ordering in the periodic Anderson model. 
The Hamiltonian is given by
\begin{align}
\label{PAM_Ham}
{\mathcal{H}}   
=& -  t \sum_{\langle i,j\rangle,\sigma} 
( c^{\dagger}_{i \sigma} c_{j \sigma}  + {\mathrm{H.c.}} ) 
-  t_f \sum_{\langle i,j\rangle,\sigma} 
( f^{\dagger}_{i \sigma} f_{j \sigma}  + {\mathrm{H.c.}} ) \nonumber \\
-& V\sum_{i ,\sigma} 
( c^{\dagger}_{i \sigma}f_{i \sigma}+{\mathrm{H.c.}} )  
+ U \sum_i n_{i \uparrow}^f n_{i \downarrow}^f + E_0 \sum_{i, \sigma} n_{i \sigma}^f, 
\end{align} 
where $f^{\dagger}_{i \sigma}$($f_{i \sigma}$) is the creation (annihilation) 
operator of ``localized" $f$ electrons with spin $\sigma$ at site $i$, and 
$n_{i\sigma}^{f} = f_{i\sigma}^\dagger f_{i\sigma}$. 
The first (second) term 
represents the 
kinetic energy of conduction $c$ (``localized" $f$) electrons, the third 
term the on-site $c$-$f$ hybridization, the fourth term the on-site 
Coulomb interaction for $f$ electrons, and the 
fifth term the atomic energy of $f$ electrons. 
The sum of $\langle i, j \rangle$ is taken over the 
nearest-neighbor sites on the cubic lattice. 
The periodic Anderson model is reduced to the Kondo lattice model in Eq.~(\ref{Kondo_Ham_RS}) in the large $U$ limit with one $f$ electron per site; 
$f$ electrons give localized moments, which couple with conduction electrons via the Kondo coupling $J \propto V^2/U$. 
We focus on the commensurate filling, 
$n^{{\rm tot}} =(1/N)\sum_{i \sigma} \langle c_{i \sigma}^{\dagger}c_{i \sigma} + f_{i \sigma}^{\dagger} f_{i \sigma}  \rangle = 1.5$, 
which corresponds to the 1/4-filling case in the Kondo lattice model~\cite{note_on_filling}.

In order to 
determine the ground state of the model in Eq.~(\ref{PAM_Ham}), 
we employ the standard Hartree-Fock 
approximation for the Coulomb $U$ term, which preserves the ${\rm SU}$(2) symmetry of the system:
We decouple $n_{i\uparrow}^{f}n_{i\downarrow}^{f}$ as 
\begin{align}
&n_{i\uparrow}^{f}n_{i\downarrow}^{f}\sim n_{i\uparrow}^{f}\langle n_{i\downarrow}^{f} \rangle
+ \langle n_{i\uparrow}^{f}\rangle n_{i\downarrow}^{f} -\langle n_{i\uparrow}^{f}\rangle \langle n_{i\downarrow}^{f}\rangle \notag  \\
&- f^{\dagger}_{i\uparrow}f_{i\downarrow} \langle f^{\dagger}_{i\downarrow}f_{i\uparrow} \rangle
- \langle f^{\dagger}_{i\uparrow}f_{i\downarrow} \rangle f^{\dagger}_{i\downarrow}f_{i\uparrow}
+ \langle f^{\dagger}_{i\uparrow}f_{i\downarrow} \rangle \langle f^{\dagger}_{i\downarrow}f_{i\uparrow}\rangle.  \notag 
\end{align}
In the calculations, we adopt 
the $2^3$-site unit cell, as shown in Fig.~\ref{Fig:ponti_Dirac_dispersion}(a). 
We confirm that the overall phase diagram 
is not altered in the calculations by the
size of the unit cell by using the $4^3$-site unit cell. 

\begin{figure}[htb!]
\begin{center}
\includegraphics[width=8cm,clip]{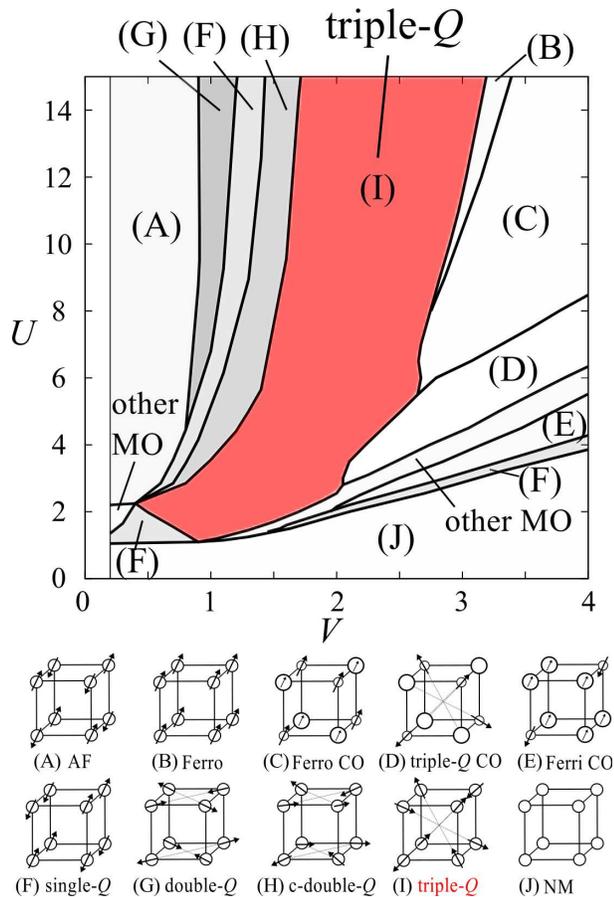} 
\caption{
(Color online). Ground-state phase diagram of 
the periodic Anderson model [Eq.~(\ref{PAM_Ham})] at 3/4 filling obtained by the mean-field calculations. 
$E_0$ and $t_f$ are taken at $-4$ and $0.2$, respectively. 
Schematic pictures of the ordering patterns in $f$ electrons are also shown. 
The sizes of circles reflect local electron densities, and the arrows represent local spin moments. 
CO stands for  charge-ordered states. 
C-double-$Q$ and other MO represent the canted double-$Q$ and other magnetically ordered states, respectively. 
}
\label{Fig:Phase_diagram}
\end{center}
\end{figure}

\begin{figure}[htb!]
\begin{center}
\includegraphics[width=5.5cm,clip]{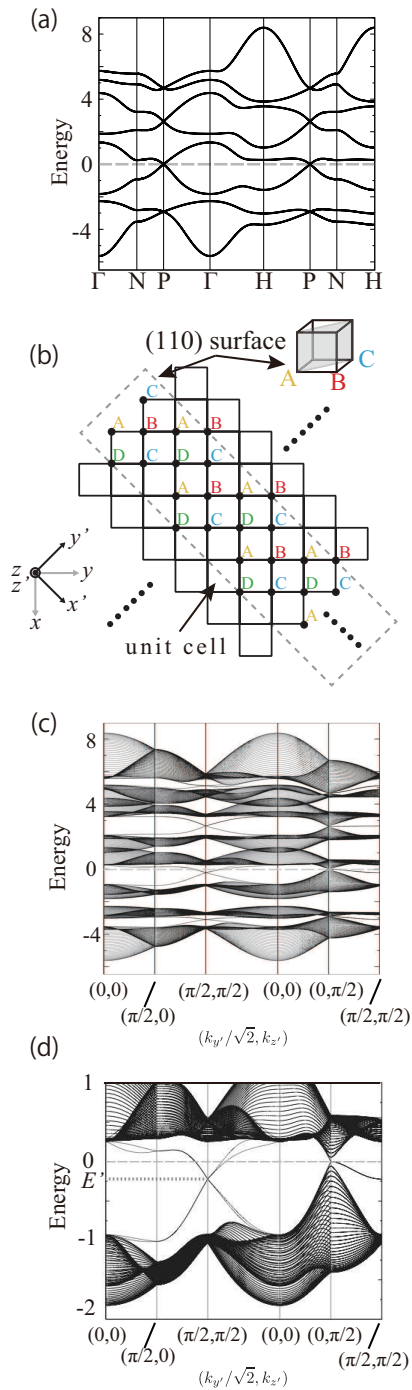} 
\caption{
(Color online). (a) Bulk electronic structure 
in the noncoplanar triple-$Q$ magnetically-ordered phase at $U=7$, $V=2.2$, $E_0=-4$, and $t_f=0.2$. 
The Fermi level at $n^{{\rm tot}}=1.5$ is set at zero. 
(b) Schematic picture of a projection of the system with the (110) surfaces onto the (001) plane.
The sites in the dashed box represent the unit cell used for the calculations of the surface states. 
(c) Energy dispersion for the system in (b) at the same parameter used in (a). 
(d) Enlarged figure of (c) in the vicinity of the Fermi level. 
}
\label{Fig:surface_dispersion}
\end{center}
\end{figure}

Figure~\ref{Fig:Phase_diagram} shows the ground-state phase diagram 
obtained by the mean-field calculations. 
Schematic pictures of magnetic and charge states for $f$ 
electrons are shown in the bottom panel of 
Fig.~\ref{Fig:Phase_diagram}. 
The result shows that various magnetic and CO states 
emerge in between the N\'{e}el-type collinear AF state in the $U\gg V$ region and 
the nonmagnetic (NM) state in the large $V$ region. 
This indicates that the model in Eq.~(\ref{PAM_Ham}) has many instabilities 
which preempt the quantum critical point 
between AF and NM phases in the 
so-called Doniach phase diagram~\cite{Doniach1977231}.

One of the dominant instabilities is a 
triple-$Q$ magnetic order, in which the $f$ spin configuration is 
equivalent to that in Eq.~(\ref{eq:tripleQ}) [Fig.~\ref{Fig:ponti_Dirac_dispersion}(a)]. 
The result strongly suggests that the triple-$Q$ state is indeed stabilized in the periodic Anderson model. 
We note that similar triple-$Q$ states were 
observed in intermetallic dysprosium compounds such as DyCu~\cite{wintenberger1971structure,Morin_1989interplay}, 
and their origin is attributed to strong magnetic anisotropy along the local [111] directions. 
Our triple-$Q$ state is further stabilized by including such magnetic anisotropy. 

As in the previous Kondo lattice case, the band structure in this phase has the 3D Dirac nodes at the $P$ point, as shown in Fig.~\ref{Fig:surface_dispersion}(a). 
In this case also, each band is doubly degenerate, while there are totally 16 bands. 
From the similar arguments in Sec.~\ref{sec:3D_Dirac}, we confirmed the emergence of essentially the same Dirac electrons as in Eq.~(\ref{effective_Weyl_Hamiltonian2}). 

Other dominant instabilities in the phase diagram in Fig.~\ref{Fig:Phase_diagram} are the CO 
insulators. 
It is noteworthy 
that a noncoplanar magnetic ordering appears in a CO state with charge density modulation at wave vector $(\pi,\pi,\pi)$. 
This is presumably due to the emergent frustration under CO; 
the charge-poor sites comprise a frustrated FCC lattice~\cite{hayami2013charge}. 
Note that, on a two-dimensional square lattice, such frustration does not appear, and the CO state is accompanied by a collinear AF order~\cite{MisawaYoshitakeMotome}.

\section{Surface Electronic structure}
\label{sec:electronic_structure}

Let us discuss the electronic state in 
the triple-$Q$ 
phase more closely, with emphasis on the peculiar surface states associated with the 3D massless Dirac states. 
For this purpose, here we take the triple-$Q$ 
magnetically-ordered 
phase (without CO) in the periodic Anderson model in Sec~\ref{sec:PAM}. 

\begin{figure}[tb!]
\begin{center}
\includegraphics[width=6cm,clip]{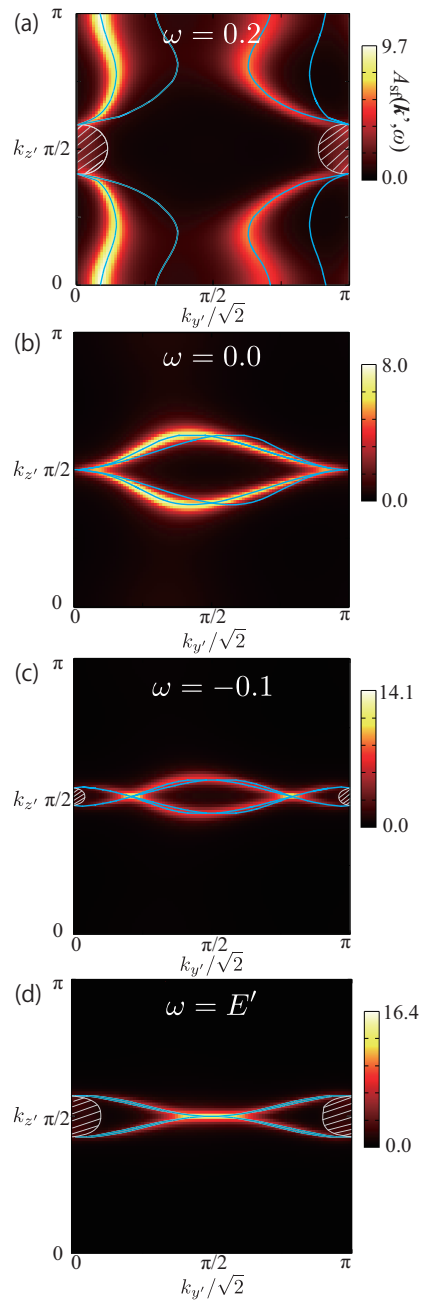} 
\caption{
(Color online). 
(a)-(d) Intensity of the single-particle spectral functions $A_{{\rm sf}} ({\bm k}', \omega)$ at $\omega=0.2$, $0.0$, $-0.1$, and $E'$, respectively. See also Fig.~\ref{Fig:surface_dispersion}(d). 
We take the broadening factor $\delta = 0.03$ in Eq.~(\ref{eq:Akw}). 
The thin lines show the constant energy contours at each $\omega$, and the hatched regions show the bulk states. 
}
\label{Fig:Fermi_Ak}
\end{center}
\end{figure}

We here consider the system with the (110) surfaces, 
in which both top and bottom surfaces consist of A and C sublattice sites [see Fig.~\ref{Fig:ponti_Dirac_dispersion}(a)]~\cite{comment_edge}; 
the geometry viewed from the $z$ direction is schematically shown in Fig.~\ref{Fig:surface_dispersion}(b). 
Figures~\ref{Fig:surface_dispersion}(c) and \ref{Fig:surface_dispersion}(d) show the band dispersions of the system with the 
(110) surfaces~\cite{comment_111}. 
The Dirac nodes at the $P$ point in the bulk system are projected onto $\bm{k}' \equiv  (k_y'/\sqrt{2}, k_z')= (0, \pi/2)$, as shown in Figs.~\ref{Fig:surface_dispersion}(c) and \ref{Fig:surface_dispersion}(d) [see Fig.~\ref{Fig:surface_dispersion}(b) for the relation between $\bm{k}$ and $\bm{k}'$]. 
Between the bulk states, there appear four bands crossing the Fermi level. 
These are the gapless surface states emergent in the triple-$Q$ state with 3D Dirac electrons. 
The four bands meet at $(\pi/2, \pi/2)$ (we set this energy $E'$), 
whereas this point is not a Dirac node as 
the band dispersion in the $(\pi/2,\pi/2)$-$(0,\pi/2)$ direction 
is not linear but quadratic in $\bm{k}'$, as shown in Fig.~\ref{Fig:surface_dispersion}(d). 

The resultant surface electronic state exhibits peculiar behavior, Fermi ``arcs" connecting the projected Dirac points. 
We note that the Fermi ``arcs" are similar to those 
found in the Weyl semimetal~\cite{Wan_PhysRevB.83.205101}. 
This is shown by calculating the single-particle spectral function at one of the two surfaces. 
The spectral function is defined as
\begin{align}
A_{{\rm sf}} 
 ({\bm k}', \omega) = -\frac{1}{\pi} {\rm Tr_{sf}} 
 [ {\rm Im}  (\omega
 + i \delta - \mathcal{H }({\bm k}'))^{-1} 
 ], 
\label{eq:Akw}
\end{align}
where the trace Tr$_{\rm sf}$ is taken only for the surface states at one of the surfaces and $\delta$ is a broadening factor. 
The result at the Fermi energy ($\omega=0$) is shown in Fig.~\ref{Fig:Fermi_Ak}(b). 
As clearly shown in the figure, 
the surface states do not
have the ordinary closed Fermi surfaces but have the Fermi ``arcs"; 
although the surface states at the Fermi level seemingly have closed forms as shown by the thin curves in the figure, 
their spectral weights become vanishingly small around the bulk Dirac cones at 
$\bm{k}' =(0, \pi/2)$ and $(\pi,\pi/2)$.  

The topology of the Fermi ``arcs" as well as Fermi surfaces, however, changes drastically while shifting the Fermi level in a rigid band picture, 
as demonstrated in Fig.~\ref{Fig:Fermi_Ak}. 
This characteristic change of the surface states in the triple-$Q$ ordered state 
is observable by the angle-resolved photoemission spectroscopy, as was recently done for Tungsten, 
in which Dirac-cone-like surface states also appear~\cite{Miyamoto_PhysRevLett.108.066808}. 
Furthermore, the spin polarization due to the surface magnetic moments will be detected in the current triple-$Q$ state. 

We note that the spectral weights of the surface states decrease rapidly by going from the 
edge to the bulk. 
The wave function at the energy $E'$ [Fig.~\ref{Fig:Fermi_Ak}(d)] 
is the most localized at the surfaces. 
On the other hand, the wave function at the Fermi level at $(0, \pi/2)$
[Fig.~\ref{Fig:Fermi_Ak}(b)] 
is extended throughout the system.
Such situations are similar to the case in zigzag-edged graphene nanoribbons~\cite{Fujita_JPSJ.65.1920,Nakada_PhysRevB.54.17954}.

\section{Summary and Concluding Remarks}
\label{sec:summary}

To summarize, we have studied the nature of 3D massless Dirac electrons on a cubic lattice, which are 
induced by a noncoplanar triple-$Q$ magnetic order. 
By using the $\boldsymbol{k}\cdot\boldsymbol{p}$ perturbation theory, 
we have shown that the low-energy excitations at particular 
commensurate fillings obey the Dirac equation. 
The Dirac state is doubly degenerate, resulting in the realization of Weyl electrons 
when the degeneracy is lifted in applied magnetic field. 
Hence, our result provides an example of Weyl semimetals on unfrustrated lattices. 
In addition, we have shown that the stability of the triple-$Q$ ordered state in two models
on the cubic lattice. 
One is the Kondo lattice model with classical localized spins. 
For this model, from the complementary studies by a variational calculation for the ground state and Monte Carlo simulation at finite temperature, we have shown that 
the triple-$Q$ ordered state appears as a stable phase in the weak-to-intermediate coupling region. 
The other model is the periodic Anderson model, for which we have also shown that the triple-$Q$ state is realized by the mean-field approximation. 
We have investigated the surface electronic state in the triple-$Q$ phase, and found that the surface states connecting the Dirac points exhibit peculiar Fermi-``arc" behavior in their spectral weights.

One of the candidate materials which exhibit 
the triple-$Q$ magnetic order with 
3D massless Dirac electrons could be DyCu, which was 
suggested to have the triple-$Q$ 
magnetic order by using neutron scattering~\cite{wintenberger1971structure,Morin_1989interplay}.
However, the electronic structure including 
the shape of Fermi surfaces is not clarified yet, to the best of our knowledge. 
Further experiments, such as the angle-resolved photoemission spectroscopy, 
are desirable to investigate the possibility of the 3D Dirac electrons.
The detailed $ab$ $initio$ band calculation for this 
material is also an intriguing future issue.

Lastly, we mention about the relationship between 
triple-$Q$ magnetic orders and 
AIII topological insulators~\cite{ryu2010topological}.
The AIII topological insulators are the 3D topological insulators
that possess a chiral symmetry, while do not possess
time-reversal and particle-hole symmetries. 
Hence, they are expected to be realized in 3D magnets in which the time-reversal symmetry is broken. 
However, its experimental realization has not been found so far, to our knowledge.
Although the 3D $\pi$-flux model 
was theoretically proposed for the AIII topological insulator~\cite{PhysRevB.81.045120},
its microscopic origin is not clear.
Our finding is crucial in this viewpoint: 
We have found that the triple-$Q$ magnetic order in the Kondo lattice model naturally
leads to the 3D $\pi$-flux model without mass term in the strong coupling limit.
By introducing perturbations giving mass term, an AIII topological insulator will be realized 
in the triple-$Q$ ordered phase.
The exploration of such possibility is left for future study.

\begin{acknowledgments}
SH and TM acknowledge Yutaka Akagi, Sho Nakosai, and Masafumi Udagawa for fruitful discussions. 
SH is supported by Grant-in-Aid for JSPS Fellows. 
Numerical calculation was partly carried out at the Supercomputer Center, Institute for Solid State Physics, University of Tokyo. 
This work was supported by Grants-in-Aid for Scientific Research (No. 23102708 and 24340076), 
the Strategic Programs for Innovative Research (SPIRE), MEXT, and the Computational Materials Science Initiative (CMSI), Japan. 
\end{acknowledgments}

\setcounter{figure}{0}
\renewcommand{\figurename}{Fig. A}

\appendix
\section{Low-energy Hamiltonian in the triple-$Q$ phase} 

\label{Low-energy Hamiltonian in the triple-Q phase}
In this Appendix, we derive the low-energy effective Hamiltonian in the triple-$Q$ phase. 
The Hamiltonian around the $P$ point is rewritten by
\begin{align}
 \tilde{\mathcal{H}} &= \tilde{\mathcal{H}}_{\bm{k}}+\tilde{\mathcal{H}}_{\bm{\Delta}}, 
\end{align}
where 
\begin{align}
 \tilde{\mathcal{H}}_{\bm{k}}&=2t (\kappa_{x} \tau_{0} \sigma_{3} + \kappa_{y} \tau_{3} \sigma_{0} + \kappa_{z} \tau_{3} \sigma_{3})  , \\
 \tilde{\mathcal{H}}_{\bm{\Delta}} &=\Delta (\tau_{0}\sigma_{1}-\tau_{2} \sigma_{2} + \tau_{2} \sigma_{3}).
\end{align}
Here, $\sigma$ and $\tau$ are the Pauli matrices, and  
$\kappa_{\alpha}=(k_{\alpha}-\pi/2)$ for $\alpha=x, y, z$. 
Let us introduce two unitary matrices $U_{1}$ and $U_2$ to diagonalize $\tilde{\mathcal{H}}_{\bm{\Delta}}$. 
$U_{1}$ and $U_2$ are defined as 
\begin{align}
U_{1}
=\begin{bmatrix} 1 & 1 \\ i  & -i \end{bmatrix}\sigma_{0}, 
\quad U_{2}&=\begin{bmatrix} 1 & 0 \\ 0  & \sigma_{1} \end{bmatrix}u,
\end{align}
respectively, where the matrix $u$ is defined as 
\begin{equation} 
u = \left( \begin{array}{cc}
-\frac{(1+i)\sqrt{\sqrt{3}-1}}{2(3)^{1/4}} & \frac{(1+i)\sqrt{\sqrt{3}+1}}{2(3)^{1/4}} \\
\frac{1}{(3)^{1/4}\sqrt{\sqrt{3}-1}} &\frac{1}{(3)^{1/4}\sqrt{\sqrt{3}+1}} \\
\end{array} \right) 
\end{equation}
By using the relations such as
\begin{align}
U_{1}^{\dagger}\tilde{\mathcal{H}}_{\bm{\Delta}}U_{1}&=\Delta(\tau_{0}\sigma_{1}-\tau_{3} \sigma_{2} + \tau_{3} \sigma_{3}), \\
u^{\dagger} (\sigma_1 - \sigma_2 + \sigma_3) u &= \sqrt{3} \sigma_3, 
\end{align}
we obtain
\begin{align}
U_{2}^{\dagger}U_{1}^{\dagger}\tilde{\mathcal{H}}_{\bm{\Delta}}U_{1}U_{2} = 
\begin{pmatrix} \sqrt{3}\Delta\sigma_{3} & 0 \\ 0 & \sqrt{3}\Delta\sigma_{3} \\ \end{pmatrix}.
\end{align}
Hence, by multiplying the unitary matrix $U=U_{1}U_{2}$ on the Hamiltonian 
$\tilde{\mathcal{H}}=\tilde{\mathcal{H}}_{\bm{k}}+\tilde{\mathcal{H}}_{\bm{\Delta}}$, we end up with 
\begin{align}
&U^{\dagger}\tilde{\mathcal{H}}U= \\ \nonumber
&\begin{pmatrix}
-2t \kappa_x u^{\dagger} \sigma_3 u + \sqrt{3} \Delta \sigma_3 & -2t u^{\dagger} (\kappa_y \sigma_0 + i \kappa_z \sigma_2) u \\
-2t u^{\dagger} (\kappa_y \sigma_0 -i \kappa_z \sigma_2)  u & 2t \kappa_x u^{\dagger} \sigma_3 u + \sqrt{3} \Delta \sigma_3 \\
\end{pmatrix}.
\end{align}
\begin{widetext}
For $\kappa_{\alpha}t/\Delta$=0, 
the spectrum of $\tilde{\mathcal{H}}$ is given as $E=\pm \sqrt{3} \Delta$, which are two doublets. 
Up to the first order of $\kappa_{\alpha}t/\Delta$,
the low-energy Hamiltonians lifting the degeneracy of these doublets 
are given by 
\begin{eqnarray}
{\tilde{\mathcal{H}}}_{+} &=&
\left( \begin{array}{cc}
\sqrt{3} \Delta - 2 t (|a|^2 -c^2) \kappa_x & -2tc [(a+ a^*) \kappa_y - (a - a^*) \kappa_z]  \\
-2tc [(a+ a^*) \kappa_y + (a - a^*) \kappa_z] & \sqrt{3} \Delta + 2 t (|a|^2 -c^2) \kappa_x \\
\end{array} \right), \\ 
{\tilde{\mathcal{H}}}_{-} &=&
\left( \begin{array}{cc}
-\sqrt{3} \Delta - 2 t (|b|^2 -d^2) \kappa_x & -2td [(b+ b^*) \kappa_y - (b - b^*) \kappa_z]  \\
-2td [(b+ b^*) \kappa_y + (b - b^*) \kappa_z] & -\sqrt{3} \Delta + 2 t (|b|^2 -d^2) \kappa_x \\
\end{array} \right).
\end{eqnarray}
\end{widetext}
Then, Eq.~(\ref{effective_Weyl_Hamiltonian2}) is obtained by rescaling $\kappa_{\alpha}$ in $\tilde{\mathcal{H}}_{+}$ and $\tilde{\mathcal{H}}_{-}$ in the form: 
\begin{align}
-(|a|^2-c^2)\kappa_x & \rightarrow \frac{1}{\sqrt{3}}\kappa_{z}, \\ 
-c(a+a^{*})\kappa_y & \rightarrow \frac{1}{\sqrt{3}} \kappa_{x}, \\ 
c(a-a^{*})\kappa_z & \rightarrow -\frac{1}{\sqrt{3}} i\kappa_{y}, 
\end{align}
for $\tilde{\mathcal{H}}_{+}$, and 
\begin{align}
-(|b|^2-d^2)\kappa_x & \rightarrow -\frac{1}{\sqrt{3}}\kappa_{z}, \\
-d(b+b^{*})\kappa_y & \rightarrow -\frac{1}{\sqrt{3}} \kappa_{x}, \\
d(b-b^{*})\kappa_z & \rightarrow \frac{1}{\sqrt{3}} i\kappa_{y},
\end{align}
for $\tilde{\mathcal{H}}_{-}$, respectively. 
Here, $a$, $b$, $c$, $d$ are the matrix elements of $u$: $u=\left( \begin{array}{cc}
a & b \\
c &d \\
\end{array} \right)$.

\section{Relation between the triple-$Q$ Hamiltonian and the 3D $\pi$-flux Hamiltonian} 
\label{Appendix2}
\begin{figure}[h!]
\begin{center}
\includegraphics[width=6cm,clip]{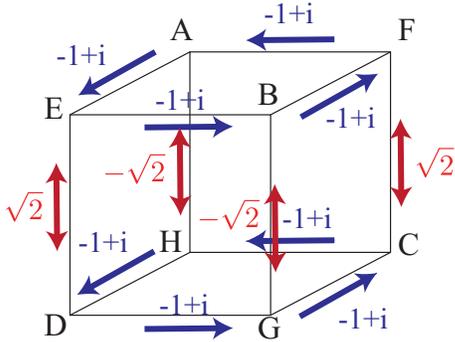} 
\caption{ 
Schematic picture of the eight-site unit cell. 
Hopping amplitudes in Eq.~(\ref{hopping_DE}) are also shown for each bond in the figure. 
}
\label{Fig:cubic}
\end{center}
\end{figure}

Here, we consider the relation between the triple-$Q$ Hamiltonian and the 
3D $\pi$-flux Hamiltonian~\cite{ryu2010topological}. 
We start from the Hamiltonian with the nearest-neighbor hoppings $-t$. 
By choosing the eight-site unit cell, the Hamiltonian is written by 
\begin{align}
\tilde{\mathcal{H}}_{{\rm hop}} = -2 t (\cos k_x \rho_{x} \tau_{0} \sigma_{x} 
+ \cos k_y \rho_{0} \tau_{0} \sigma_{x} + \cos k_z  \rho_{x} \tau_{x} \sigma_{x}), 
\end{align} 
where $\rho$, $\tau$, and $\sigma$ are the Pauli matrices. 
The basis are represented by 
\begin{eqnarray}
(c_{A \bm{k}}, c_{B \bm{k}}, c_{C \bm{k}}, c_{D \bm{k}}, c_{E \bm{k}}, c_{F \bm{k}}, c_{G \bm{k}}, c_{H \bm{k}}), 
\end{eqnarray}
where the capital subscript represents the sites, as shown in Fig.~\ref{Fig:cubic}. 
Now, we introduce the effect of the exchange coupling to the 
triple-$Q$ magnetic order in the strong coupling limit ($\Delta/t \rightarrow \infty$) as shown in the second term in Eq.~(\ref{Kondo_Ham_kspace}). 
Then, the hopping amplitude $t$ is modified as 
\begin{align}
 t \rightarrow t \cos  \frac{\theta_{i}}{2}  \cos \frac{\theta_{j}}{2} 
 + \sin \frac{\theta_{i}}{2}  \sin \frac{\theta_{j}}{2}  {\rm e}^{-i (\phi_{i}-\phi_{j})},
 \label{hopping_DE}
\end{align}
where $i$ and $j$ are site index, and we have introduced the polar coordinates [$\bm{S}_{i} \propto (\sin \theta_{i} \cos \phi_{i}, \sin \theta_i \sin \phi_i, \cos \theta_i )$] for the direction of the local triple-$Q$ magnetic field. 
In that case, the triple-$Q$ Hamiltonian in the strong coupling limit is represented by 
\begin{eqnarray}
\tilde{\mathcal{H}}_{{\rm hop}}^{{\rm DE}} =&-&\sqrt{\frac{2}{3}} t [\cos k_x( -\rho_{x} \tau_{0} \sigma_{x}+ \rho_{x} \tau_{z} \sigma_{y}) \\ \nonumber
&-& \cos k_y (\rho_{0} \tau_{0} \sigma_{x}+\rho_{0} \tau_{z} \sigma_{y}) 
+\sqrt{2} \cos k_z  \rho_{x} \tau_{y} \sigma_{y}]. 
\label{DE_Ham}
\end{eqnarray}
The energy spectrum of $\tilde{\mathcal{H}}_{{\rm hop}}^{{\rm DE}}$ is given by 
\begin{eqnarray}
E_{\bm{k}}= \pm \frac{2}{\sqrt{3}} t \sqrt{\cos^2 k_x + \cos^2 k_y + \cos^2 k_z}. 
\end{eqnarray}
There are four degeneracy for each band and the Hamiltonian in Eq.~(\ref{DE_Ham}) has Dirac dispersions around the $(\pi/2,\pi/2, \pi/2)$ point. 

Furthermore, we found that the triple-$Q$ Hamiltonian in the strong coupling limit is equivalent to the 3D $\pi$-flux 
Hamiltonian in Ref.~\onlinecite{ryu2010topological} by multiplying the unitary matrix $U_3$: 
\begin{align}
\tilde{\mathcal{H}}_{\pi {\rm flux} } &= \sqrt{3} U_3^{\dagger} \mathcal{H}_{{\rm hop}}^{{\rm DE}} U_3 \\
&= -2 t ( \cos k_x  \rho_{x} \tau_{0} \sigma_{x}+ \cos k_y \rho_{z} \tau_{z} \sigma_{x}
-  \cos k_z \rho_{x} \tau_{x} \sigma_{y}).  
\end{align}
$U_3$ is defined by 
\begin{eqnarray}
U_3=
\left(
\begin{array}{cc}
\tilde{\bm{U}}_3 & \tilde{\bm{O}} \\
\tilde{\bm{O}} & \tilde{\bm{I}}
\end{array}
\right),
\end{eqnarray}
where $\tilde{\bm{O}}$ and $\tilde{\bm{I}}$ are the $4 \times 4$ null matrix and identity matrix, respectively.  
$\tilde{\bm{U}}_3$ is represented by
\begin{widetext}
\begin{eqnarray}
\frac{1}{c^2}\left(
\begin{array}{cccc}
 e^{i \frac{\pi}{4}} c_x^2-e^{-i \frac{\pi}{4}}  c_y^2+ c_z^2 
 &  \sqrt{2} c_x c_y 
 & (1-e^{i \frac{\pi}{4}} ) c_x c_z 
 & (-1-e^{-i \frac{\pi}{4}}) c_y c_z  \\
  - \sqrt{2} i c_x c_y 
 &  e^{i \frac{\pi}{4}} c_x^2+e^{-i \frac{\pi}{4}} c_y^2+ c_z^2 
 & ( 1-e^{-i \frac{\pi}{4}} ) c_y c_z 
 & ( 1-e^{i \frac{\pi}{4}} ) c_x c_z  \\
 (-1+e^{-i \frac{\pi}{4}}) c_x c_z
 & (-1+e^{i \frac{\pi}{4}} ) c_y c_z 
 & e^{-i \frac{\pi}{4}} c_x^2+e^{i \frac{\pi}{4}} c_y^2+c_z^2
 &  \sqrt{2} i c_x c_y  \\
 (1+e^{i \frac{\pi}{4}} ) c_y c_z 
 &  (-1+e^{-i \frac{\pi}{4}} ) c_x c_z 
 &  \sqrt{2} c_x c_y 
 & e^{-i \frac{\pi}{4}} c_x^2-e^{i \frac{\pi}{4}} c_y^2+ c_z^2  \\
\end{array}
\right),  
\end{eqnarray}
\end{widetext}
where we describe $\cos k_x \rightarrow c_x, \cos k_y \rightarrow c_y, \cos k_z \rightarrow c_z$ 
for simplicity, and $c^2=c_x^2+c_y^2+c_z^2$.  


\begin{thebibliography}{32}
\bibitem[{com({\natexlab{a}})}]{comment_Sq}
\bibinfo{note}{We also checked the correlation of the spin structure factor
  $\langle S_{\bm{Q}_{1}}S_{\bm{Q}_{2}}S_{\bm{Q}_{3}} \rangle$ to distinguish
  the triple-$Q$ state from the double-$Q$ and single-$Q$ states.}
\bibitem[{not()}]{note_on_filling}
\bibinfo{note}{We confirm that a similar triple-$Q$ state is also stabilized at
  $n=5/2$.}
\bibitem[{com({\natexlab{b}})}]{comment_edge}
\bibinfo{note}{When we cut one of the surfaces at B and D sublattice sites, a
  small gap opens at the Dirac cones. The gap magnitude decreases while
  increasing the system size along the $x'$ direction and disappears in the
  bulk limit.}
\end{thebibliography}
\bibliographystyle{apsrev}

\end{document}